\documentclass[12pt,twoside]{article}
\usepackage{fleqn,espcrc1}

\usepackage{graphicx}
\usepackage[figuresright]{rotating}

\newcommand{\AmS}{{\protect\the\textfont2
  A\kern-.1667em\lower.5ex\hbox{M}\kern-.125emS}}

\title{
\vspace{-39mm}
\hfill {\small Preprint DFPD 00/TH/29, TRI-PP-00-28}
\vspace{34mm}\\
\bf Spin observables for pion production 
from $pd$ collisions}

\author{L. Canton\address{INFN and University of Padova, 
        via F. Marzolo n. 8, 35131 Padova, Italy}$^{\rm ,c}$,
        G. Pisent$^{\rm a}$,
        W. Schadow\address{TRIUMF, 4004 Wesbrook Mall, 
        Vancouver, B.C., V6T 2A3  Canada},
and     J. P. Svenne\address{Physics Department, University of Manitoba, 
        Winnipeg, MB, R3T 2N2 Canada}  }

\begin{document}

\maketitle

\begin{abstract}
We have calculated the proton analyzing power $A_{y0}$ of the
pion-production reaction from $pd$ collisions for one energy close to
threshold and for another in the region of the $\Delta$-resonance.  A
fair reproduction of the experimental data could be obtained in both
cases with a model which includes isoscalar and isovector $\pi N$
rescatterings in $s$ waves, as well as the $p$-wave rescattering
mechanisms mediated by the $\pi NN$ and $\pi N \Delta$ vertices. For
the analyzing power at threshold we found that the initial-state
interaction (ISI) is also quite important.

\end{abstract}

\section{INTRODUCTION}

Pion-production reactions have the potential to probe the nuclear
phenomena at short distance since they typically involve processes
transferring large momenta to the target nucleus.  On the other hand,
the pion is also the principal mediator of the nuclear force, hence,
meson production (or absorption) plays a fundamental role in hadron
dynamics and the study of these processes may reveal aspects of the
meson-baryon coupling, and more generally of meson-exchange phenomena,
which would remain otherwise hidden.

Reliable calculations on meson dynamics in few-nucleon systems are
extremely difficult, and usually one has to assume that various
reaction mechanisms compete in dominating the process. The
description of the process is complicated because the treatment of the
reaction mechanisms reveals the occurrence of many terms, and this
forces one to make further assumptions in order to reduce the number
of terms to a few, tractable ones. This reduction clearly introduces
ambiguities making it more difficult to extract information about the
nuclear wave function at short distances, or about the modifications of
the hadron interactions because of the presence of other nucleons.  A
somewhat simplified situation is found when considering
nucleon-induced production close to the pion threshold, since there
the $s$-wave mechanisms of the elementary $NN$$\rightarrow\pi$$NN$
inelasticities dominate, while the $p$-wave mechanisms (including the
isobar degrees of freedom) can be treated as corrections.

\section{THEORETICAL MODEL}

In a recent paper~\cite{1}, the $\pi$-production reaction from $pd$
collisions has been calculated in the threshold region. The model
includes meson-exchange rescattering diagrams in $s$ waves, with $\pi
N$ contributions in the isovector and the isoscalar channels. The
isovector contribution was generated by a $\rho$-mediated diagram, and
the isoscalar interaction by means of a subtracted $\sigma$-exchange
model which is suppressed on-shell but enhanced when the rescattered
pion is off-mass shell~\cite{Hamilton}.  This off-shell enhancement
has been previously advocated~\cite{Oset} to describe the production
cross section at threshold for the simpler reaction $pp\rightarrow
pp\pi^o$ and represents an alternative explanation for this process
with respect to the one given by Lee and Riska~\cite{Lee} in terms of
heavy-meson exchange (HME) effects.  With the same parameterization
suggested by Hamilton we found in Ref.~\cite{1} that the
pion-production cross section at threshold can be reproduced, while
without these off-shell effects the results are underestimated by one
order of magnitude.

According to a previous treatment~\cite{2} centered around
the $\Delta$-resonance region, the effects of the $\pi N$ $p$-wave 
interaction have also been included, via the nonrelativistic 
$\pi NN$ vertex, as well as by means of the $\pi N\Delta$ interaction. 
The intermediate $\Delta$ propagation terminates through a 
$\Delta N$ transition determined by  the $\pi$ plus  $\rho$ exchange 
diagrams, where the pseudoscalar meson provides the typical longitudinal 
structure to the transition potential, while the vector meson generates 
the transverse contribution. 

\section{RESULTS}

With the same model of Ref.~\cite{1}, we have calculated the spin
observables at threshold. The formalism for the spin observables for
this reaction has been given in Ref.~\cite{5}. For the 3$N$
bound-state in the outgoing channel, we have used the wave functions
which have been calculated in Ref. \cite{3}. As two-nucleon input for
the three-nucleon equations we have used high rank separable
representations \cite{Haidpriv} of the Paris and the Bonn {\em B}
potentials, known as PEST and BBEST, respectively.  These
representations were originally constructed by the Graz group~\cite{4}.
The three-nucleon dynamics in the initial state (ISI) has been
calculated by solving the Faddeev-type Alt-Grassberger-Sandhas
equations~\cite{Alt}.  More details can be found in Refs.~\cite{1,2}.

\begin{figure}[htb]
\begin{minipage}[t]{80mm}
\includegraphics[width=5.5cm,angle=-90]{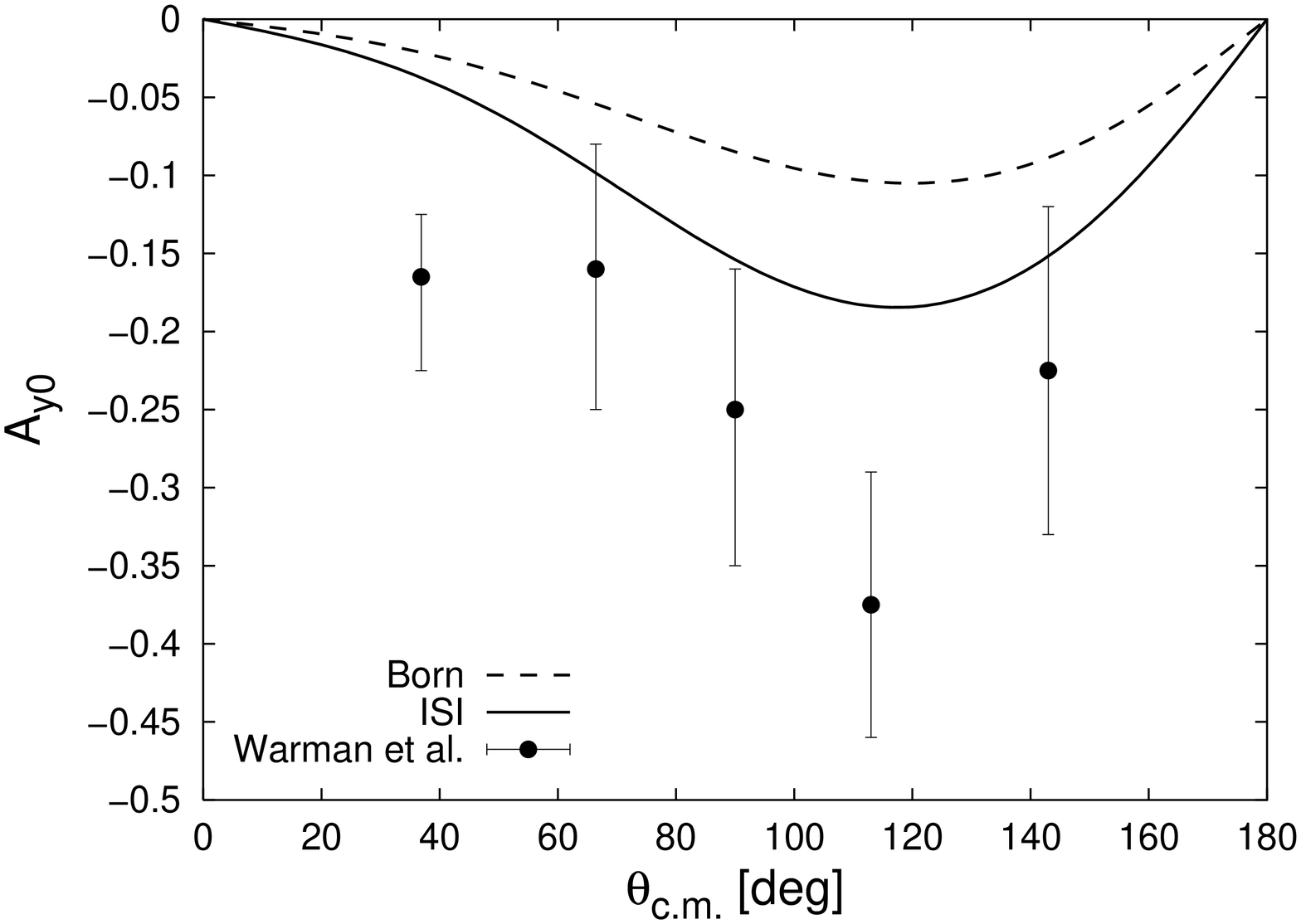}
\caption{Proton analyzing power $A_{y0}$ for the 
$\vec{p}d\rightarrow\pi^o {\ }^3$He process close to threshold.
Both calculations include all discussed mechanisms, and 
the solid line includes also the effect of 3$N$ initial-state interaction,
while the dashed line is the result obtained in Born approximation.  
The data are from Ref.~\cite{6}.}
\label{fig:threshold}
\end{minipage}
\hspace{\fill}
\begin{minipage}[t]{75mm}
\hspace{-4mm}
\includegraphics[width=5.5cm,angle=-90]{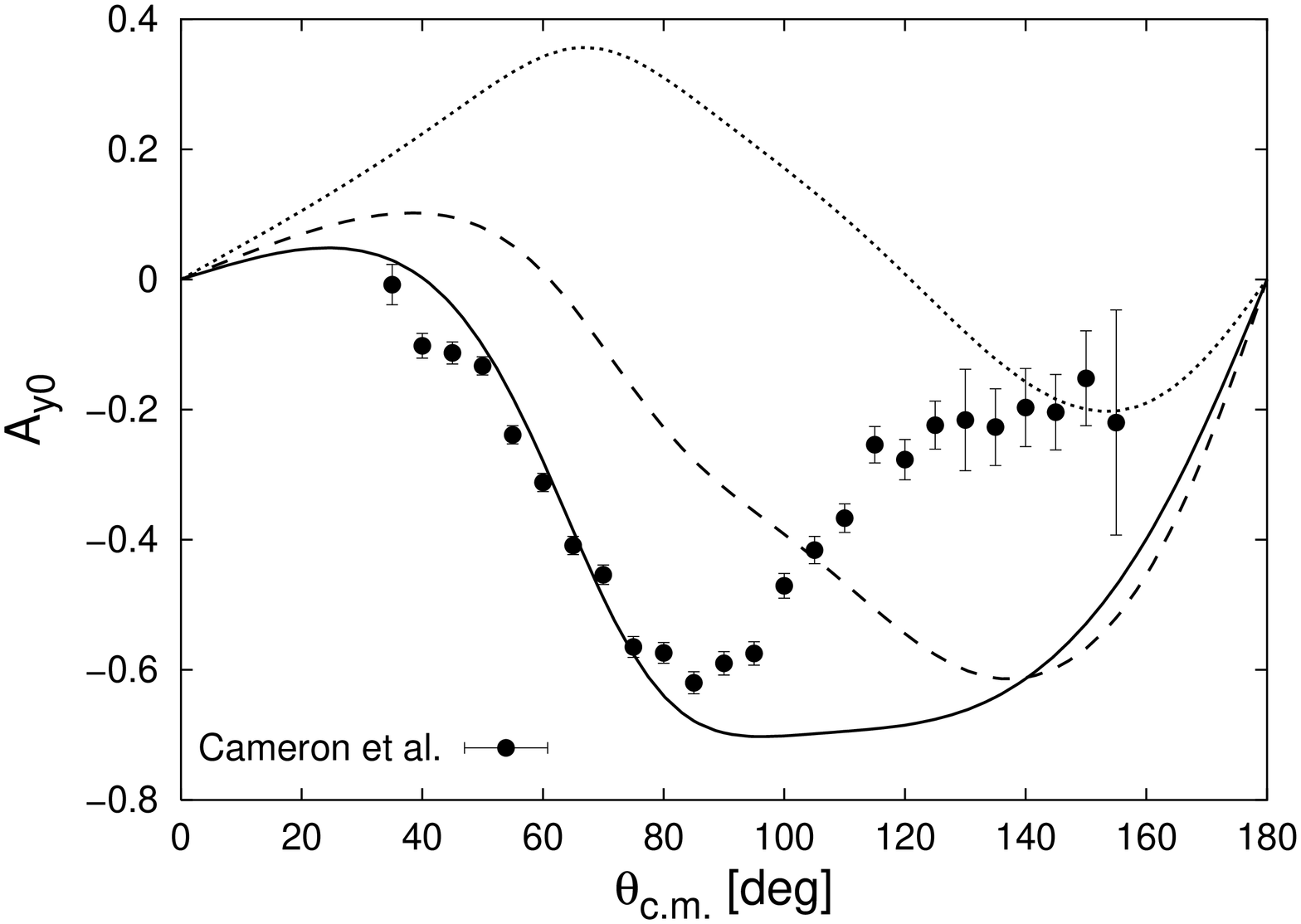}
\caption{$A_{y0}$ for energies
around the $\Delta$ resonance. The solid line includes
isoscalar off-shell effects, isovector $\rho$ exchanges, 
$\Delta$ rescatterings, and the $\pi$$NN$ vertex.
The results when excluding the isoscalar off-shell effects are 
given by the dashed line.
The dotted line contains only the $p$-wave mechanisms.
The data are from Ref.~\cite{7}.}
\label{fig:deltares}
\end{minipage}
\end{figure}

As an example, the angular distribution for the proton analyzing power
$A_{y0}$ is shown for the pion center-of-mass momentum $P_{c.m.}$
$\simeq 0.25 \, m_\pi$, close to the $\pi$-production threshold
(Fig.~\ref{fig:threshold}), and also at higher energy ($P_{c.m.}\simeq
1.36 \, m_\pi$), close to the $\Delta$ resonance
(Fig.~\ref{fig:deltares}).  The two lines shown in
Fig.~\ref{fig:threshold} include all production mechanisms we have
discussed, and differ because the solid line includes also the effects
of ISI, while the dotted line refers to a plane-wave calculation. The
ISI effects turn out to be quite important, while at higher energy
their effects are smaller.  In Fig.~\ref{fig:deltares} we show the
modifications obtained when adding the two $s$-wave diagrams
(``$\rho$'' and ``$\sigma$'') on top of the $p$-wave mechanisms, at
the $\Delta$ resonance. Both these two diagrams are needed for the
reproduction of $A_{y0}$ in the forward hemisphere, while at backward
angles the situation is still controversial.

\hspace{4mm} 

{\bf Acknowledgments}

L. C. and G. P. acknowledge funds from the ``Ministero
dell'Universit\'a e della Ricerca Scientifica e Tecnologica'', within
the Project ``Fisica Teorica del Nucleo e dei Sistemi a Pi\`u Corpi''.
W. Sch. and J. P. S. acknowledge support and hospitality from the Physics
Department of the University of Padova, and from INFN, sez. di
Padova. The work of W. Sch. and J.~P.~S. is also supported by the
Natural Science and Engineering Research Council of Canada.

\end{document}